\documentclass{PoS}

\usepackage{amsmath}
\usepackage{amssymb}
\usepackage{cite}
\usepackage{subfigure}

\def\d{\hbox{d}}
\def\gosam{\texttt{GoSam}}
\def\sherpa{\texttt{Sherpa}}
\def\samurai{\texttt{Samurai}}
\def\ninja{\texttt{Ninja}}
\def\golem{\texttt{Golem95}}
\def\form{\texttt{FORM}}
\def\qgraf{\texttt{QGRAF}}
\def\haggies{\texttt{Haggies}}
\def\spinney{\texttt{Spinney}}
\def\powheg{\texttt{POWHEG BOX}}
\def\amcatnlo{\texttt{aMC@NLO}}
\def\herwig{\texttt{HERWIG++}}
\def\MINLO{\texttt{MiNLO}}
\def\PYTHIA{\texttt{PYTHIA}}

\def\HW{\textrm HW}
\def\HWJ{\textrm HWJ}

\def\done{\rm d}
\def\NNLOPS{NNLO+PS}

\def\MCatNLO{\texttt{MC@NLO}}

\title{Interfacing GoSam with Monte Carlo Event Generators}

\ShortTitle{GoSam interfaced to MC generators}

\author{Gionata Luisoni\thanks{In collaboration with G.~Cullen, H.~Van~Deurzen, N.~Greiner, G.~Heinrich, P.~Mastrolia, E.~Mirabella, G.~Ossola, T.~Peraro,
J.~Reichel, J.~Schlenk, J.F.G.~von~Soden-Fraunhofen, F.~Tramontano and P.~Nason, C.~Oleari, S.~Höche, J.~Huang, M.~Schönherr, J.~Winter.}\\
        Max-Planck-Institut f\"ur Physik, F\"ohringer Ring 6,\\D-80805 M\"unchen, Germany\\
        E-mail: \email{luisonig@mpp.mpg.de}}

\abstract{In this talk the most recent results obtained by interfacing \gosam{} with external Monte Carlo event generators are presented and summarized. In the last year the automatic one-loop amplitude generator \gosam{} has been used for the computation of several processes relevant for the LHC physics program. In the first part of the talk the latest results are summarized and the status of the interfaces to several external Monte Carlo programs, based on the Binoth-Les-Houches-Accord, is reported. The second part is dedicated to two selected computations. One concerning the associated production of a Higgs and a vector boson in association with $0$ and $1$ jet computed with \gosam{}+\powheg{}, and one focusing on the analysis of the forward--backward asymmetry in the production of top quark pairs using $0$ and $1$ jet merged samples with \gosam{}+\sherpa{}. Finally some recent results on Beyond-Standard-Model (BSM) physics are also presented.}

\FullConference{11th International Symposium on Radiative Corrections (Applications of Quantum Field Theory to Phenomenology) (RADCOR 2013),\\
		22-27 September 2013\\
		Lumley Castle Hotel, Durham, UK }

\begin{document}

\section{Introduction}
In the recent past the developments in the computation of Next-to-Leading Order (NLO) corrections to Standard Model (SM) processes allowed to automatize all the different parts of the computation leading to the development of tools for the automatic calculation of NLO processes with Monte Carlo event generators. A NLO cross section can be canonically written as the sum of different ingredients
\begin{align}
\sigma_{\rm NLO}=&\int_{\d\Phi_{m}}\d\sigma_{\rm Born}
+\int_{\d\Phi_{m}}\left[\d\sigma_{\rm NLO}^{V}+{\int_{\d\Phi_{1}}}\d\sigma_{\rm NLO}^{\rm S}\right]
+\int_{\d\Phi_{m+1}}\left(\d\sigma_{\rm NLO}^{\rm R}-\d\sigma_{\rm NLO}^{\rm S}\right)\,,
\end{align}
where the first term on the right hand side gives the Leading Order (LO) contribution, the second term the NLO virtual contribution and the integrated subtraction terms used to subtract the divergences from the real radiation part, which is encoded in the last term. Many MC generators can nowadays generate automatically tree-level amplitudes for Born and real contributions, recognizing the singular limits of the latter and constructing the corresponding subtraction terms. The computation of the virtual contribution instead is usually left to dedicated programs. In the last years many tools were developed, which allow to automatically compute one-loop Quantum Chromodynamic~(QCD) and Electroweak~(EW) amplitudes for arbitrary SM processes with up to $4,5$ final-state particles~\cite{Cullen:2011ac,Badger:2010nx,Hirschi:2011pa,Bevilacqua:2011xh,Cascioli:2011va,Actis:2012qn,Bern:2013pya}. In the following we will present some recent phenomenological applications of the automatic one-loop amplitude generator \gosam{}~\cite{Cullen:2011ac}.

\section{\gosam{}}
The \gosam{} program is a framework for the automatic generation and numerical computation of one-loop amplitudes. The core program consists of a python package which generates fortran95 code for the evaluation of the desired one-loop amplitudes. The amplitudes are based upon the algebraic generation of $d$-dimensional integrands via Feynman diagrams, which can be evaluated both using integrand-reduction techniques and tensor integral calculus. This approach implies also the possibility to generate and compute on-the-fly the full rational term, without the need of further ad-hoc Feynman rules. For the generation of the diagrams we use \qgraf{}~\cite{Nogueira:1991ex}, whereas for the algebraic manipulation of the raw amplitudes expressions we are committed to \form{}~\cite{Kuipers:2012rf} and the package \spinney{}~\cite{Cullen:2010jv}. Finally the algebraic expressions are converted to optimized fortran95 code using \form{} and \haggies{}~\cite{Reiter:2009ts}.

At running time the diagrams can be evaluated using \samurai{}~\cite{Mastrolia:2010nb}, which performs a reduction at the integrand level~\cite{Ossola:2006us} or with \golem{}~\cite{Cullen:2011kv}, which is a library for the computation of one-loop tensor integrals. A new reduction method, based on the Laurent expansion of the integrand was developed recently and implemented into a new code called \ninja{}~\cite{Mastrolia:2012bu}. The reduction method can be changed at running time. In the computations presented in the next sections \samurai{} is used as the default program and \golem{} is used as rescue program whenever an instability is detected in the integrand-reduction approach.

This approach allows one to easily compute one-loop QCD and EW corrections to processes within the SM and also beyond, by simply using an appropriate model file. While for QCD corrections the whole computation is fully automatized, for EW corrections and BSM physics the user has to provide, by hand, the correct renormalization.

\section{NLO computations with \gosam{} and the BLHA}
For the computation of full NLO processes \gosam{} is equipped with an interface using the Binoth-Les-Houches Accord (BLHA) standards~\cite{Binoth:2010xt}. This allows an automatic interface with external MC programs. Recently an update of this interface was published~\cite{Alioli:2013nda}, with the aim of increasing its automation and flexibility. Among the new features there is the possibility to pass dynamical parameters like couplings and masses among the MC and the One-Loop-Program (OLP). Furthermore the synchronization of EW schemes is now simplified and standards for the treatment of unstable phase-space points as well as for merging different jet multiplicities were set. Driven by the need of some MCs the interface was also extended to be able to exchange color- and helicity-correlated tree amplitudes.

So far three different MC event generators were successfully interfaced to \gosam{} to compute NLO cross sections and distributions for a number of different processes, summarized in Table~\ref{tab:interface}.

An ad-hoc setup was used to interface with the MadGraph4-MadEvent-MadDipole~\cite{Stelzer:1994ta} family of programs. It was used to compute NLO QCD corrections to the production of $b\bar{b}b\bar{b}$~\cite{Greiner:2011mp} and later to assess the impact of the still missing one-loop contributions in the production of $W^{+}W^{-}\,jj$~\cite{Greiner:2012im}, both at hadron colliders. The $b\bar{b}b\bar{b}$-production process was recently computed considering massive b-quarks~\cite{Bevilacqua:2013taa}, whereas the latter was first computed in~\cite{Melia:2011dw}, where parts of the loop contributions were neglected. More recently the same setup was used to compute NLO QCD corrections to some BSM processes~\cite{Cullen:2012eh,Greiner:2013gca}, whose results are presented more extensively in Section~\ref{sec:bsm}, and to the production of a photon pair in association with one and two jets~\cite{Gehrmann:2013aga}. Finally \gosam{} was also used to compute the production of a pair of Higgs bosons in association with 2 jets in~\cite{Dolan:2013rja}.

Using the BLHA interface \gosam{} was successfully interfaced with the \powheg{}~\cite{Frixione:2007vw}. This new interface, together with the built-in interface to MadGraph4 allows for a quick generation of new processes in the \powheg{} framework. In Section~\ref{sec:powheg} we report more details about the computation of the associated production of a Higgs, a vector boson and a jet~\cite{Luisoni:2013cuh}.

Thanks to the BLHA interface it was possible to link in a easy and automatic way also to the MC event generator \sherpa{}~\cite{Gleisberg:2007md}. The two programs can exchange information about the process the user wishes to compute and generate all the needed ingredients. The user only needs to fill an input card for the two programs and, once the code is ready, the full calculation can be steered simply editing the \sherpa{} input card. A set of ready-to-use process packages with the full code for the loop computation is also available~\cite{Gosam:web}. They only require the installation of \sherpa{} by the user, whereas the code for the virtual part is already generated and validated. This setup was also used to compute the production of a Higgs boson in association with 2 jets in gluon-gluon fusion~\cite{vanDeurzen:2013rv} and the production of a Higgs boson with a top-antitop pair and a jet~\cite{vanDeurzen:2013xla}. More details about these computations can be found in these proceedings as well~\cite{pierpaoloproc}. Furthermore the \gosam{}+\sherpa{} setup was used to perform a MC analysis of the $t\bar{t}$ forward-backward asymmetry $A_{\rm FB}$~\cite{Hoeche:2013mua} to which Section~\ref{sec:afb} is dedicated. Finally a further analysis of $A_{\rm FB}$ with off-shell top-quarks is in preparation and was also presented at this conference~\cite{johannesproc}.

A hybrid setup making use of the best features of both the MadGraph4-MadEvent-MadDipole family and Sherpa was used together with \gosam{} to compute the NLO QCD corrections to the production of a Higgs boson is association with three jets~\cite{Cullen:2013saa}. We refer again to a parallel contribution~\cite{pierpaoloproc} to these proceedings for more details.

Interfaces to \amcatnlo{}~\cite{aMCatNLO:web} and \herwig{}~\cite{Bellm:2013lba} are currently under development.

\begin{table}
\begin{center}
\begin{tabular}{|l l|}
\hline
\multicolumn{2}{|c|}{\textbf{GoSam + MadGraph4-MadEvent-MadDipole}}\\
\hline
$pp\to b\bar{b} b\bar{b}$\,\cite{Greiner:2011mp} & $pp\to \tilde{\chi}_1^0 \tilde{\chi}_1^0$+jet \,\cite{Cullen:2012eh}\\
$pp\to \gamma\gamma$+1,2\,jets \,\cite{Gehrmann:2013aga} &  $pp\to $ graviton($\to \gamma\gamma$)+jet \,\cite{Greiner:2013gca}\\
$pp\to W^+W^-$+2\,jets\,\cite{Greiner:2012im} & \\
\hline
\hline
\multicolumn{2}{|c|}{\textbf{GoSam + Sherpa}}\\
\hline
$pp\to W^+W^+ +$2\,jets & $pp\to W^\pm b\bar{b}$ (massive b-quarks)\\
$pp\to W^+W^-\,b\bar{b}$\,\cite{johannesproc} & $pp\to W^+W^-$\\
$pp\to W^\pm$+0,1,2,3 jets & $pp\to Z/\gamma$+0,1,2 jets\\
$pp\to t\bar{t}$+0,1 jets\,\cite{Hoeche:2013mua} & $pp\to t\bar{t}H$+0,1 jet\,\cite{vanDeurzen:2013xla}\\
$pp\to H+$2\,jets (gluon fusion)\,\cite{vanDeurzen:2013rv} & \\
\hline
\hline
\multicolumn{2}{|c|}{\textbf{GoSam + MadGraph4-MadEvent-MadDipole + Sherpa}}\\
\hline
$pp\to H+$3\,jets (gluon fusion) \, \cite{Cullen:2013saa} &\\
\hline\hline
\multicolumn{2}{|c|}{\textbf{GoSam + POWHEG BOX}}\\
\hline
$pp\to HW/HZ$+0,1 jet\,\cite{Luisoni:2013cuh} & \\
\hline
\hline
\multicolumn{2}{|c|}{\textbf{GoSam}}\\
\hline
$pp\to H\,H+2$\,jets\,\cite{Dolan:2013rja} & \\
\hline
\end{tabular}
\end{center}
\caption{NLO calculations done by interfacing \gosam{} with different Monte Carlo programs.}
\label{tab:interface}
\end{table}

\section{Associated production of a Higgs boson, vector boson and a jet}
\label{sec:powheg}
\begin{figure}[htb]
\begin{center}
\includegraphics[width=0.49\textwidth]{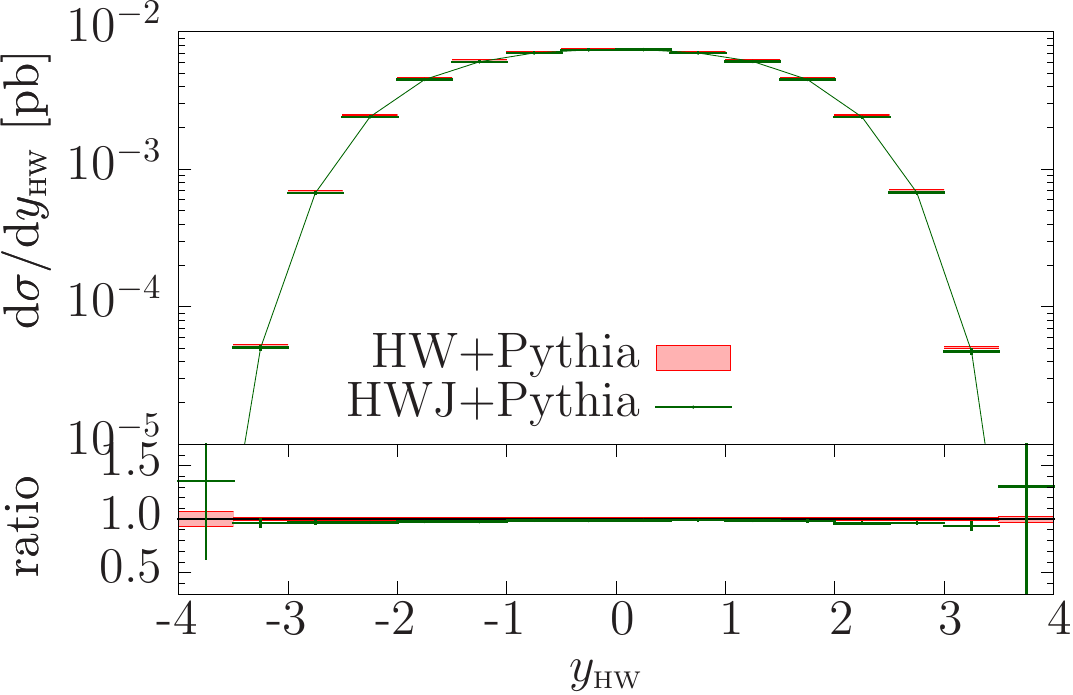}
\includegraphics[width=0.49\textwidth]{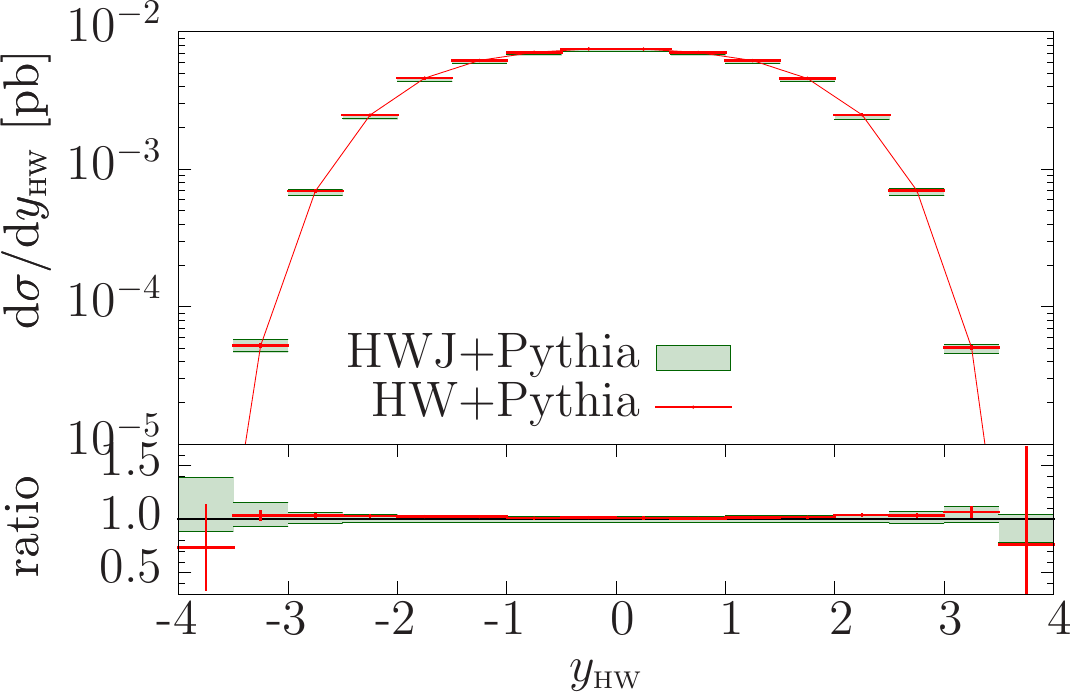}
\end{center}
\caption{Comparison between the \HW{} + \PYTHIA{} result and the
  \HWJ-\MINLO{} + \PYTHIA{} result for the $HW^-$ rapidity distribution at
  the LHC at 8~TeV.  The left plot shows the 7-point scale-variation band for
  the \HW{} generator, while the right plot shows the \HWJ-\MINLO{} 7-point
  band.}
\label{fig:HWy}
\end{figure}

\begin{figure}[htb]
\begin{center}
\includegraphics[width=0.49\textwidth]{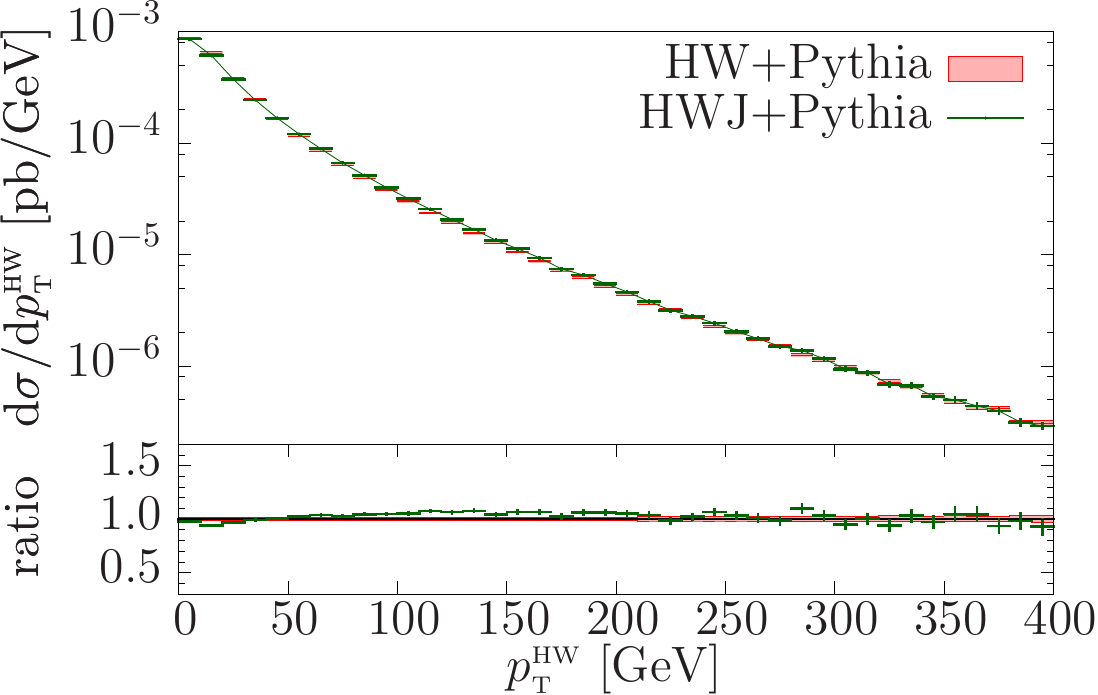}
\includegraphics[width=0.49\textwidth]{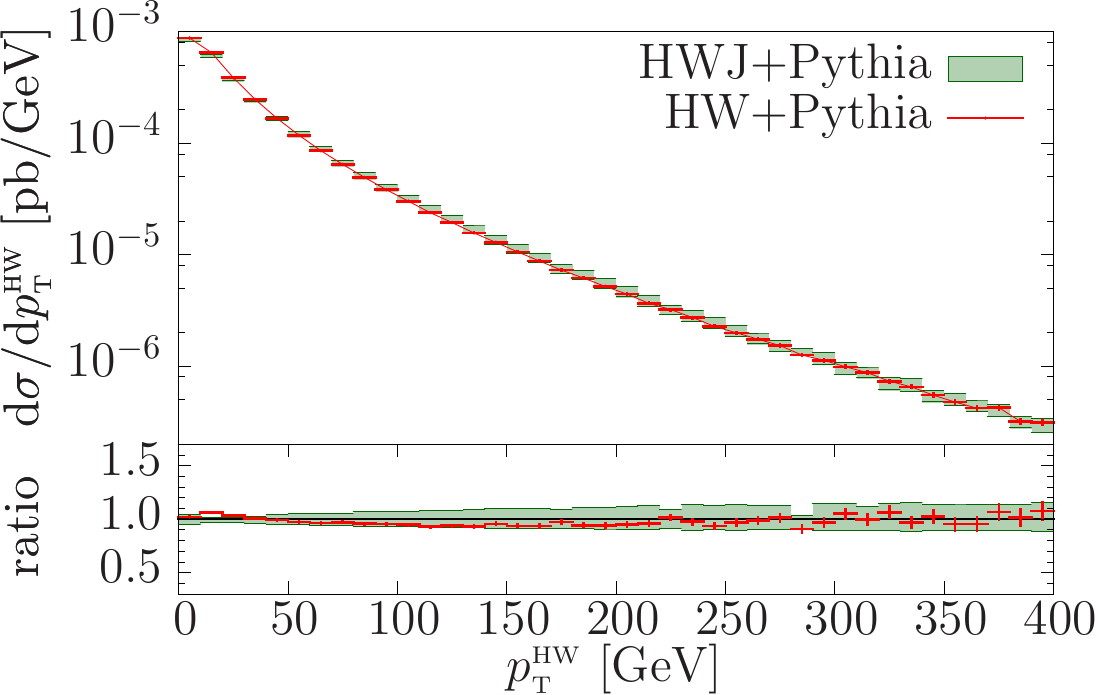}
\end{center}
\caption{Comparison between the \HW{} + \PYTHIA{} result and the
  \HWJ{}-\MINLO{} + \PYTHIA{} result for the $HW^-$ transverse-momentum
  distribution.}
\label{fig:HWpt}
\end{figure}

Higgs boson production in association with a vector boson is an interesting channel for Higgs boson studies at the LHC, since it seems to be the only available channel to study the Higgs branching to $b\bar{b}$, or to set limits to the Higgs branching into invisible particles. In~\cite{Luisoni:2013cuh} a \powheg{} generator for the computation of $HV$ and $HV$ + 1 jet (where $V=Z,W^{\pm}$) was presented. The virtual amplitudes were generated with \gosam{} and the two codes were interfaced using the BLHA interface.
Using the $HV$ + 1 jet generators with the improved version of the \MINLO{} procedure~\cite{Hamilton:2012np} it was possible to reach NLO accuracy for quantities that are inclusive in the production of the color-neutral system, i.e.~when the associated jet is not resolved. This is shown in Figures~\ref{fig:HWy}-\ref{fig:HWpt} for increasingly exclusive observables. The uncertainty bands are obtained by varying separately renormalization and factorization scales by factors of $\frac{1}{2}$ and $2$, taking thus the envelope among 7 different curves.

Whereas the agreement between the \HW{} and the \HWJ{}-\MINLO{} generators are very good for the $HW^-$ rapidity distribution (Fig.~\ref{fig:HWy}), there are small differences in the transverse momentum distribution of the $HW^-$ system (Fig.~\ref{fig:HWpt}). The reason for these little differences is due to the fact that this distribution is only computed at leading order by the \HW{} generator, while it is computed at NLO accuracy by the \HWJ{}-\MINLO{} generator. Conclusions similar to those for $HW(j)$ can be drawn for $HZ(j)$ associated production. Using the \powheg{} generator for the computation of $HV$ and $HV$ + 1 jet presented here it is furthermore possible to construct an \NNLOPS{} generator using a similar approach to the one of~\cite{Hamilton:2013fea}.

\section{Top-antitop forward-backward asymmetry analysis}
\label{sec:afb}
The forward--backward asymmetry in the production of top quark pairs offers great opportunities to study the physics both within and beyond the SM. At $p\bar{p}$ colliders, the asymmetry in dependence on the observable $O$ is defined as
\begin{equation}
  A_\mathrm{FB}(O)=\frac{\done\sigma_{t\bar{t}}/\done O|_{\Delta y>0}
    -\done\sigma_{t\bar{t}}/\done O|_{\Delta y<0}}{
    \done\sigma_{t\bar{t}}/\done O|_{\Delta y>0}
    +\done\sigma_{t\bar{t}}/\done O|_{\Delta y<0}}\;,
\end{equation}
where $\Delta y=y_t-y_{\bar t}$ is the rapidity difference between the top and the antitop quark. It was pointed out in~\cite{Skands:2012mm} that color flows from incoming quarks to the top quark and from antiquarks to the antitop quark lead to more radiation when the top quark goes backward. This generates a positive asymmetry already at the level of parton showers that include color coherence effects. The \gosam{}+\sherpa{} interface was used to generate a code providing a merged simulation of $t\bar{t}$ and $t\bar{t}+$jet production at hadron colliders, which preserves both the NLO accuracy of the fixed-order prediction and the logarithmic accuracy of the parton shower~\cite{Hoeche:2013mua}. The latter is based on Catani--Seymour dipole factorization~\cite{Schumann:2007mg} and the related \MCatNLO{} generator~\cite{Hoeche:2011fd} to generate events at the parton shower level. The two samples are merged at a scale of $7$ GeV according to the procedure explained in~\cite{Gehrmann:2012yg}.

\begin{figure}[htb]
\subfigure[]{\includegraphics[width=7.cm]{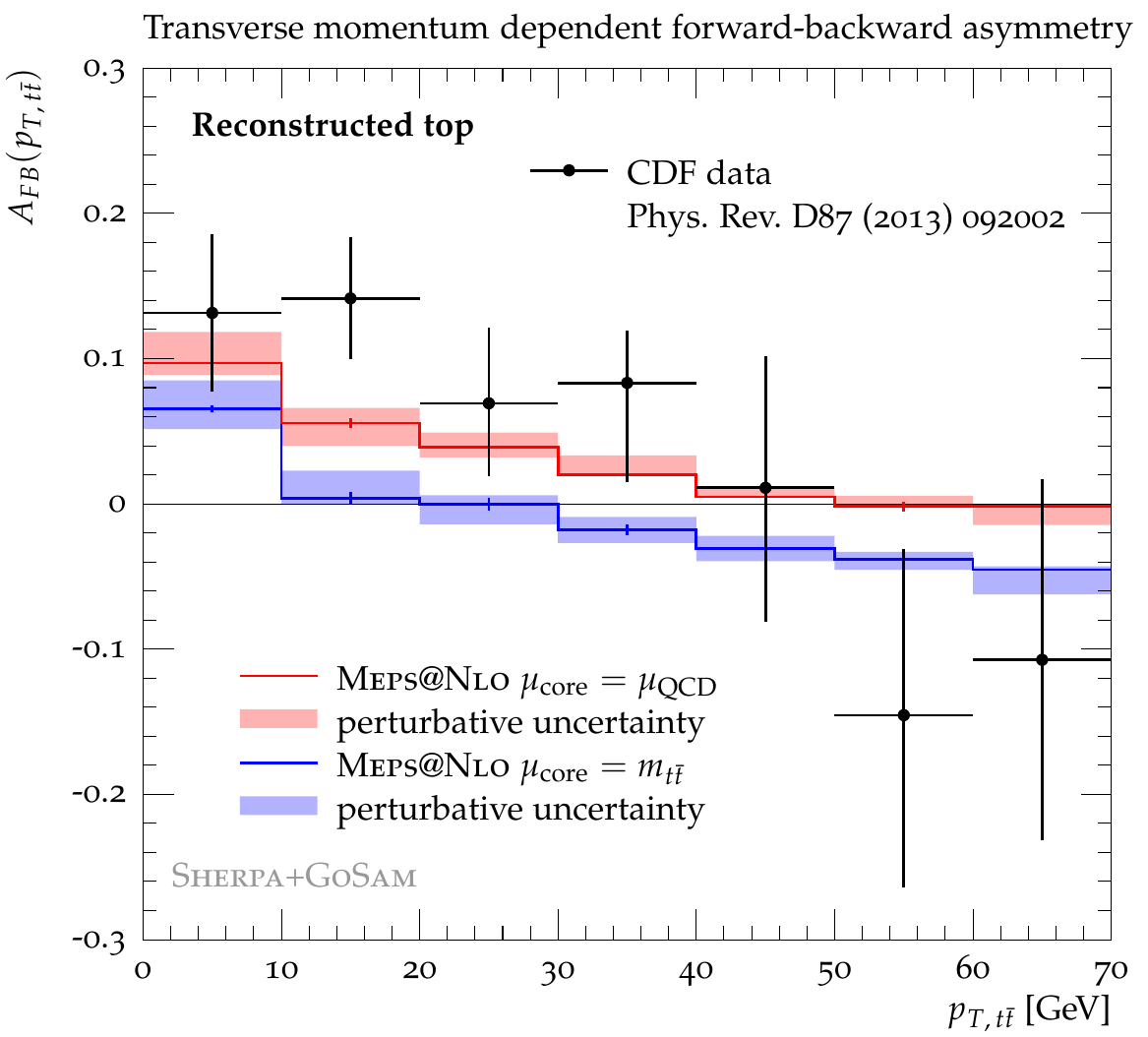}}\hfill
\subfigure[]{\includegraphics[width=7.cm]{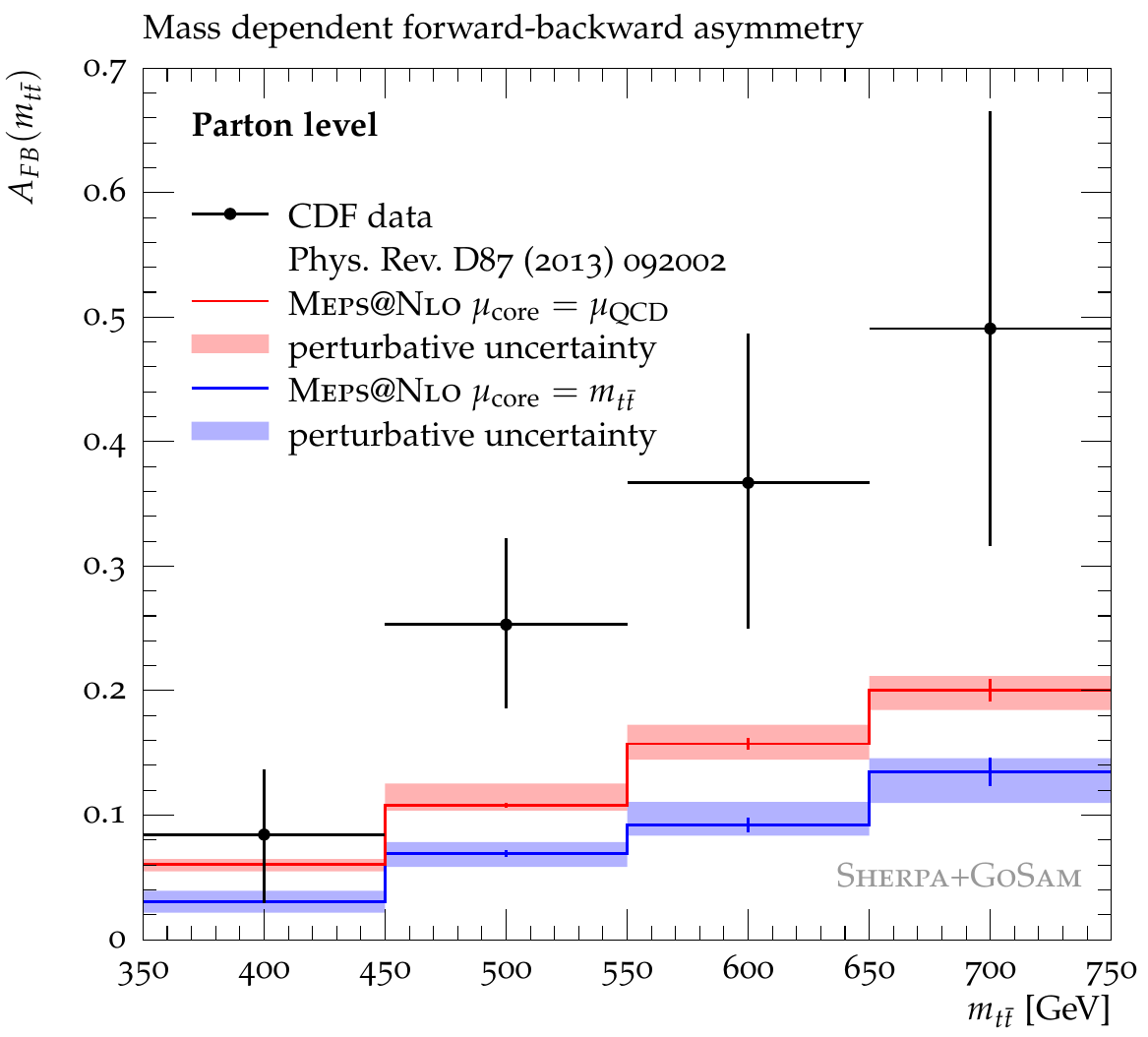}}
\caption{Top quark forward--backward asymmetry in dependence on the transverse
  momentum (a) and the invariant mass (b) of the $t\bar t$ system. \MCatNLO{} zero
  plus one jet merged predictions, together with their uncertainty bands,
  are shown for two scale choices studied in~\cite{Hamilton:2013fea}.
  The comparison is against CDF background subtracted data (a) and against
  parton-level corrected data (b)~\cite{Aaltonen:2012it}.}
\label{fig:afb}
\end{figure}

Figures~\ref{fig:afb}(a)-\ref{fig:afb}(b) show results for $A_{\rm FB}$ as a function of the transverse momentum and the invariant mass. The agreement with the CDF data for $A_\mathrm{FB}(p_{T,t\bar t})$ is good, whereas for the invariant mass the MC predictions can reproduce the linear rise but remain below the data. The former result implies a quantitative
agreement of the simulation with data in two very different phase-space domains, driven by different physics phenomena: multiple soft and virtual parton emission in the so-called Sudakov region, and hard parton radiation for larger pair transverse momenta. The disagreement in the invariant mass distribution instead can be explained by the fact that for this observable the Sudakov region is spread out over the entire range of the measurement, leading to an increase of $A_{\rm FB}$ for larger values of $m_{t\bar t}$.

\section{Beyond SM physics with \gosam{}}
\label{sec:bsm}
\begin{figure}[htb]
\subfigure[]{\includegraphics[width=8.cm]{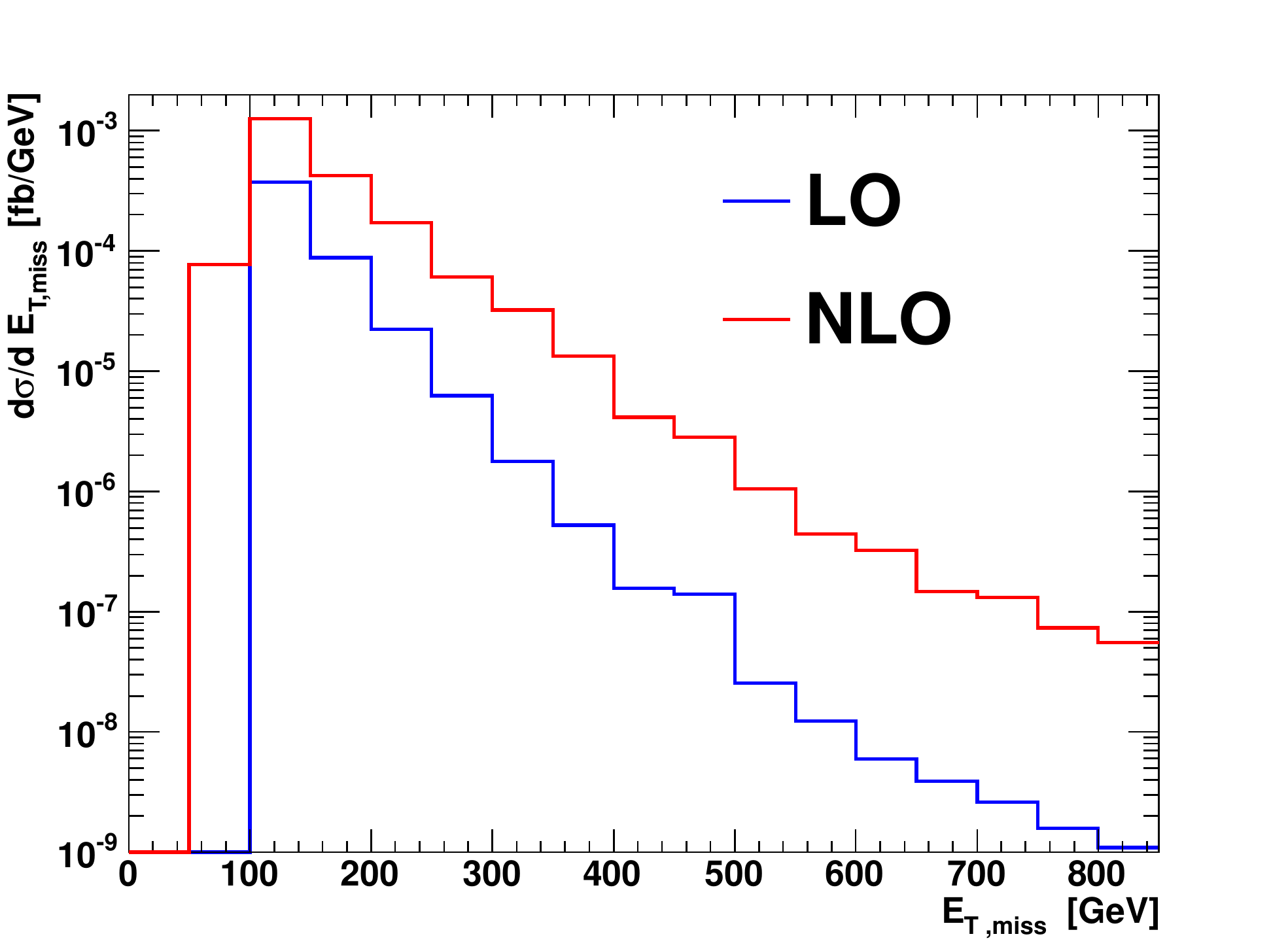}}\hfill
\subfigure[]{\includegraphics[width=7.cm]{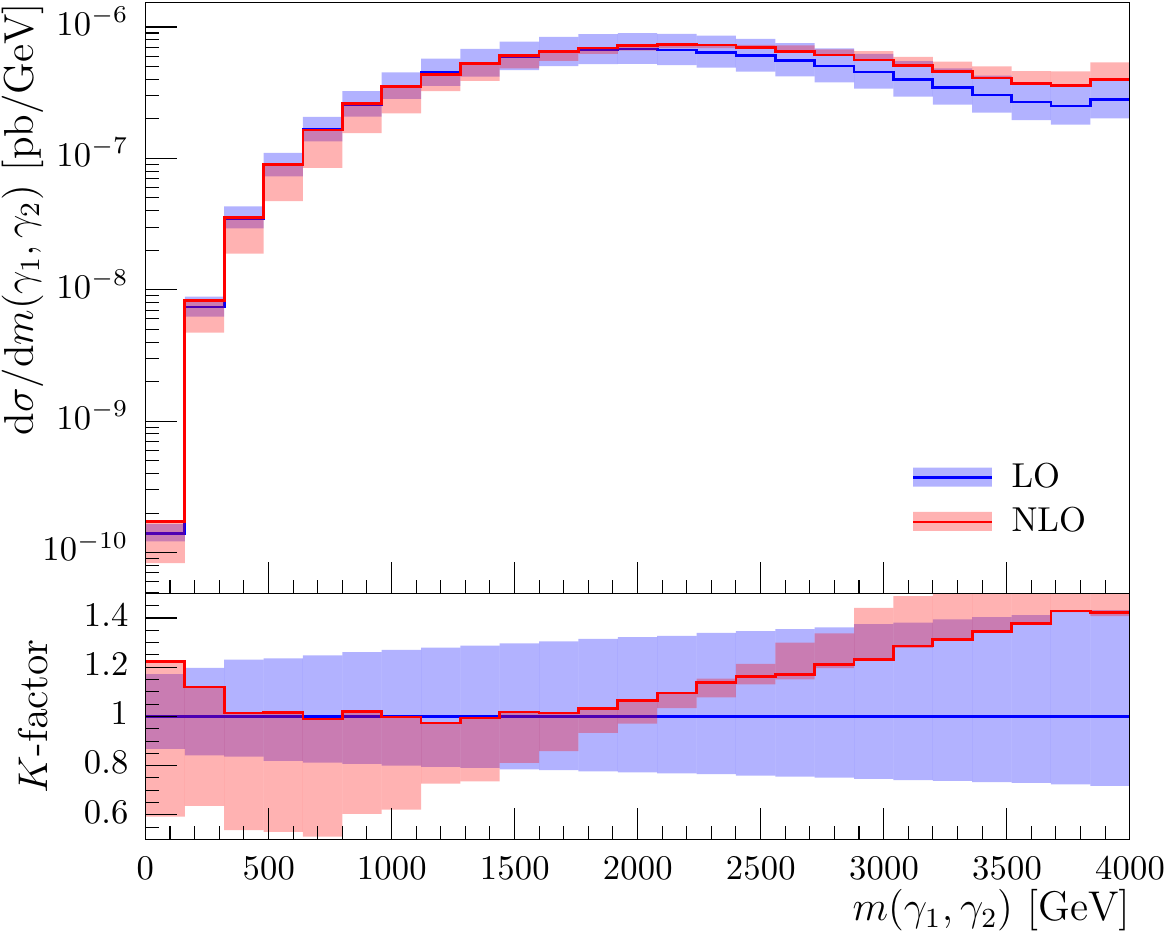}}
\caption{(a) missing transverse energy $E_T^{\rm{miss}}$ for the process $pp\to \tilde{\chi}_1^0\tilde{\chi}_1^0$+jet at $\sqrt{s}=8$\,TeV. (b) NLO QCD corrections to the invariant mass distribution of the photon pair stemming from graviton decay within the ADD model for $\delta=4$ large extra dimensions. The bands show the scale variations by a factor of two around the central scale $\mu_0^2 = \mu_F^2 = \frac{1}{4} \left(  m_{\gamma\gamma}^2 +  p_{T,jet}^2  \right)$.}
\label{fig:bsm}
\end{figure}

\gosam{} also has been used to calculate the NLO Susy-QCD corrections to the production of a pair of the lightest neutralinos plus one jet at the LHC at $8$\,TeV,
appearing as a monojet signature in combination with missing energy. All non-resonant diagrams were fully included, i.e. no simplifying assumption that production and decay
factorize were used. The resulting missing transverse energy distribution is shown in Fig.~\ref{fig:bsm}(a) and it can be seen that the NLO corrections are large, mainly due to additional channels opening up at NLO. The detailed setup can be found in~\cite{Cullen:2012eh}.

Another BSM calculation based on \gosam{} + MadGraph4-MadEvent-MadDipole are the NLO QCD corrections to the production of a graviton in association with one jet~\cite{Greiner:2013gca}, where the  graviton decays into a pair of photons, within ADD models of large extra dimensions~\cite{ArkaniHamed:1998rs}.
The calculation is quite involved due to the complicated tensor structure introduced by spin-2 particles, and the non-standard propagator of the graviton, coming from the summation over Kaluza-Klein modes. As can be seen from Fig.~\ref{fig:bsm}b, the K-factors turn out {\it not} to be uniform over the range of the diphoton invariant mass distribution. As the latter in general is used to derive exclusion limits, the differential NLO corrections should be taken into account. More details can be found in~\cite{Greiner:2013gca}.

\section*{Acknowledgments}
I would like to thank all my collaborators G.Cullen, H. Van Deurzen, N. Greiner, G.Heinrich, P.Mastrolia, E.Mirabella, G.Ossola, T.Peraro,
J. Reichel, J. Schlenk, J.F.G. von Soden-Fraunhofen and F. Tramontano working on \gosam{}, and also S.Höche, J. Huang, P. Nason, C. Oleari, M.Schönherr, J.Winter for innumerable many interesting and fruitful discussions. This work is supported by the Alexander von Humboldt Foundation, in the framework of the Sofja Kovaleskaja Award Project ``Advanced Mathematical Methods for Particle Physics'', endowed by the German Federal Ministry of Education and Research.


\begin{thebibliography}{99}
\bibitem{Cullen:2011ac}
  G.~Cullen, N.~Greiner, G.~Heinrich, G.~Luisoni, P.~Mastrolia, G.~Ossola, T.~Reiter and F.~Tramontano,
  Eur.\ Phys.\ J.\ C {\bf 72} (2012) 1889
  [arXiv:1111.2034].

\bibitem{Bern:2013pya}
  Z.~Bern, L.~J.~Dixon, F.~F.~Cordero, S.~Hoeche, H.~Ita, D.~A.~Kosower, D.~Maitre and K.~J.~Ozeren,
  [arXiv:1310.2808].

\bibitem{Badger:2010nx}
  S.~Badger, B.~Biedermann and P.~Uwer,
  Comput.\ Phys.\ Commun.\  {\bf 182} (2011) 1674
  [arXiv:1011.2900].

\bibitem{Hirschi:2011pa}
  V.~Hirschi, R.~Frederix, S.~Frixione, M.~V.~Garzelli, F.~Maltoni and R.~Pittau,
  JHEP {\bf 1105} (2011) 044
  [arXiv:1103.0621].

\bibitem{Bevilacqua:2011xh}
  G.~Bevilacqua, M.~Czakon, M.~V.~Garzelli, A.~van Hameren, A.~Kardos, C.~G.~Papadopoulos, R.~Pittau and M.~Worek,
  Comput.\ Phys.\ Commun.\  {\bf 184} (2013) 986
  [arXiv:1110.1499].

\bibitem{Cascioli:2011va}
  F.~Cascioli, P.~Maierhofer and S.~Pozzorini,
  Phys.\ Rev.\ Lett.\  {\bf 108} (2012) 111601
  [arXiv:1111.5206].

\bibitem{Actis:2012qn}
  S.~Actis, A.~Denner, L.~Hofer, A.~Scharf and S.~Uccirati,
  JHEP {\bf 1304} (2013) 037
  [arXiv:1211.6316].

\bibitem{Nogueira:1991ex}
  P.~Nogueira,
  J.\ Comput.\ Phys.\  {\bf 105} (1993) 279.

\bibitem{Kuipers:2012rf}
  J.~Kuipers, T.~Ueda, J.~A.~M.~Vermaseren and J.~Vollinga,
  Comput.\ Phys.\ Commun.\  {\bf 184} (2013) 1453
  [arXiv:1203.6543].

\bibitem{Cullen:2010jv}
  G.~Cullen, M.~Koch-Janusz and T.~Reiter,
  Comput.\ Phys.\ Commun.\  {\bf 182} (2011) 2368
  [arXiv:1008.0803].

\bibitem{Reiter:2009ts}
  T.~Reiter,
  Comput.\ Phys.\ Commun.\  {\bf 181} (2010) 1301
  [arXiv:0907.3714].

\bibitem{Mastrolia:2010nb}
  P.~Mastrolia, G.~Ossola, T.~Reiter and F.~Tramontano,
  JHEP {\bf 1008} (2010) 080
  [arXiv:1006.0710].

\bibitem{Ossola:2006us}
  G.~Ossola, C.~G.~Papadopoulos and R.~Pittau,
  Nucl.\ Phys.\ B {\bf 763} (2007) 147
  [hep-ph/0609007];
  R.~K.~Ellis, W.~T.~Giele and Z.~Kunszt,
  JHEP {\bf 0803} (2008) 003
  [arXiv:0708.2398].

\bibitem{Cullen:2011kv}
  G.~Cullen, J.P.~Guillet, G.~Heinrich, T.~Kleinschmidt, E.~Pilon, T.~Reiter and M.~Rodgers,
  Comput.\ Phys.\ Commun.\  {\bf 182} (2011) 2276
  [arXiv:1101.5595];\\
  T.~Binoth, J.P.~Guillet, G.~Heinrich, E.~Pilon and T.~Reiter,
  Comput.\ Phys.\ Commun.\  {\bf 180} (2009) 2317
  [arXiv:0810.0992].

\bibitem{Mastrolia:2012bu}
  P.~Mastrolia, E.~Mirabella and T.~Peraro,
  JHEP {\bf 1206} (2012) 095
   [Erratum-ibid.\  {\bf 1211} (2012) 128]
  [arXiv:1203.0291];

\bibitem{tizianoproc}
  T.~Peraro et al., these proceedings.

\bibitem{Binoth:2010xt}
  T.~Binoth {\it et al.},
  Comput.\ Phys.\ Commun.\  {\bf 181} (2010) 1612
  [arXiv:1001.1307].

\bibitem{Alioli:2013nda}
  S.~Alioli {\it et al.},
  Comput.\ Phys.\ Commun.\  {\bf 185} (2014) 560
  [arXiv:1308.3462].

\bibitem{Stelzer:1994ta}
  T.~Stelzer and W.~F.~Long,
  Comput.\ Phys.\ Commun.\  {\bf 81} (1994) 357
  [hep-ph/9401258];\\
  F.~Maltoni and T.~Stelzer,
  JHEP {\bf 0302} (2003) 027
  [hep-ph/0208156];\\
  J.~Alwall, P.~Demin, S.~de Visscher, R.~Frederix, M.~Herquet, F.~Maltoni, T.~Plehn and D.~L.~Rainwater {\it et al.},
  JHEP {\bf 0709} (2007) 028
  [arXiv:0706.2334];\\
  R.~Frederix, T.~Gehrmann and N.~Greiner,
  JHEP {\bf 0809} (2008) 122
  [arXiv:0808.2128];\\
  R.~Frederix, T.~Gehrmann and N.~Greiner,
  JHEP {\bf 1006} (2010) 086
  [arXiv:1004.2905].

\bibitem{Gehrmann:2010ry}
  T.~Gehrmann and N.~Greiner,
  JHEP {\bf 1012} (2010) 050
  [arXiv:1011.0321].

\bibitem{Greiner:2011mp}
  T.~Binoth, N.~Greiner, A.~Guffanti, J.~Reuter, J.~-P.~.Guillet and T.~Reiter,
  Phys.\ Lett.\ B {\bf 685} (2010) 293
  [arXiv:0910.4379];\\
  N.~Greiner, A.~Guffanti, T.~Reiter and J.~Reuter,
  Phys.\ Rev.\ Lett.\  {\bf 107} (2011) 102002
  [arXiv:1105.3624].

\bibitem{Greiner:2012im}
  N.~Greiner, G.~Heinrich, P.~Mastrolia, G.~Ossola, T.~Reiter and F.~Tramontano,
  Phys.\ Lett.\ B {\bf 713} (2012) 277
  [arXiv:1202.6004].

\bibitem{Bevilacqua:2013taa}
  G.~Bevilacqua, M.~Czakon, M.~Krämer, M.~Kubocz and M.~Worek,
  JHEP {\bf 1307} (2013) 095
  [arXiv:1304.6860].

\bibitem{Melia:2011dw}
  T.~Melia, K.~Melnikov, R.~Rontsch and G.~Zanderighi,
  Phys.\ Rev.\ D {\bf 83} (2011) 114043
  [arXiv:1104.2327].

\bibitem{Cullen:2012eh}
  G.~Cullen, N.~Greiner and G.~Heinrich,
  Eur.\ Phys.\ J.\ C {\bf 73} (2013) 2388
  [arXiv:1212.5154].

\bibitem{Greiner:2013gca}
  N.~Greiner, G.~Heinrich, J.~Reichel and J.~F.~von Soden-Fraunhofen,
  JHEP {\bf 1311} (2013) 028
  [arXiv:1308.2194].

\bibitem{Gehrmann:2013aga}
  T.~Gehrmann, N.~Greiner and G.~Heinrich,
  JHEP {\bf 06} (2013) 58
  [arXiv:1303.0824];\\
  T.~Gehrmann, N.~Greiner and G.~Heinrich,
  [arXiv:1308.3660];\\
  T.~Gehrmann, N.~Greiner and G.~Heinrich, these proceedings;
  [arXiv:1311.4754].

\bibitem{Dolan:2013rja}
  M.~J.~Dolan, C.~Englert, N.~Greiner and M.~Spannowsky,
  [arXiv:1310.1084].

\bibitem{Frixione:2007vw}
  S.~Frixione, P.~Nason and C.~Oleari,
  JHEP {\bf 0711} (2007) 070
  [arXiv:0709.2092];\\
  S.~Alioli, P.~Nason, C.~Oleari and E.~Re,
  JHEP {\bf 1006} (2010) 043
  [arXiv:1002.2581].

\bibitem{Luisoni:2013cuh}
  G.~Luisoni, P.~Nason, C.~Oleari and F.~Tramontano,
  JHEP {\bf 1310} (2013) 083
  [arXiv:1306.2542].


\bibitem{Gleisberg:2007md}
  T.~Gleisberg and F.~Krauss,
  Eur.\ Phys.\ J.\ C {\bf 53} (2008) 501
  [arXiv:0709.2881];\\
  T.~Gleisberg, S.~.Hoeche, F.~Krauss, M.~Schonherr, S.~Schumann, F.~Siegert and J.~Winter,
  JHEP {\bf 0902} (2009) 007
  [arXiv:0811.4622];\\

\bibitem{Gosam:web}
  \texttt{http://gosam.hepforge.org/}

\bibitem{vanDeurzen:2013rv}
  H.~van Deurzen, {\it et al.},
  Phys.\ Lett.\ B {\bf 721} (2013) 74
  [arXiv:1301.0493].

\bibitem{vanDeurzen:2013xla}
  H.~van Deurzen, G.~Luisoni, P.~Mastrolia, E.~Mirabella, G.~Ossola and T.~Peraro,
  Phys.\ Rev.\ Lett.\  {\bf 111} (2013) 171801
  [arXiv:1307.8437].

\bibitem{Hoeche:2013mua}
  S.~Hoeche, J.~Huang, G.~Luisoni, M.~Schoenherr and J.~Winter,
  Phys.\ Rev.\ D {\bf 88} (2013) 014040
  [arXiv:1306.2703].

\bibitem{johannesproc}
  J.~Schlenk et al., these proceedings.

\bibitem{Cullen:2013saa}
  G.~Cullen, H.~van Deurzen, N.~Greiner, G.~Luisoni, P.~Mastrolia, E.~Mirabella, G.~Ossola and T.~Peraro {\it et al.},
  Phys.\ Rev.\ Lett.\  {\bf 111} (2013) 131801
  [arXiv:1307.4737].

\bibitem{pierpaoloproc}
  P.~Mastrolia et al., these proceedings.

\bibitem{aMCatNLO:web}
  \texttt{http://amcatnlo.web.cern.ch/amcatnlo/}

\bibitem{Bellm:2013lba}
  J.~Bellm, S.~Gieseke, D.~Grellscheid, A.~Papaefstathiou, S.~Platzer, P.~Richardson, C.~Rohr and T.~Schuh {\it et al.},
  [arXiv:1310.6877].

\bibitem{Hamilton:2012np}
  K.~Hamilton, P.~Nason and G.~Zanderighi,
  JHEP {\bf 1210} (2012) 155
  [arXiv:1206.3572];\\
  K.~Hamilton, P.~Nason, C.~Oleari and G.~Zanderighi,
  JHEP {\bf 1305} (2013) 082
  [arXiv:1212.4504].

\bibitem{Hamilton:2013fea}
  K.~Hamilton, P.~Nason, E.~Re and G.~Zanderighi,
  [arXiv:1309.0017].

\bibitem{Skands:2012mm}
  P.~Skands, B.~Webber and J.~Winter,
  JHEP {\bf 1207} (2012) 151
  [arXiv:1205.1466].

\bibitem{Schumann:2007mg}
  S.~Schumann and F.~Krauss,
  JHEP {\bf 0803} (2008) 038
  [arXiv:0709.1027];\\
  S.~Hoeche, S.~Schumann and F.~Siegert,
  Phys.\ Rev.\ D {\bf 81} (2010) 034026
  [arXiv:0912.3501].

\bibitem{Hoeche:2011fd}
  S.~Hoeche, F.~Krauss, M.~Schonherr and F.~Siegert,
  JHEP {\bf 1209} (2012) 049
  [arXiv:1111.1220];\\
  S.~Hoeche and M.~Schonherr,
  Phys.\ Rev.\ D {\bf 86} (2012) 094042
  [arXiv:1208.2815].

\bibitem{Gehrmann:2012yg}
  T.~Gehrmann, S.~Hoche, F.~Krauss, M.~Schonherr and F.~Siegert,
  JHEP {\bf 1301} (2013) 144
  [arXiv:1207.5031];\\
  S.~Hoeche, F.~Krauss, M.~Schonherr and F.~Siegert,
  JHEP {\bf 1304} (2013) 027
  [arXiv:1207.5030].

\bibitem{Aaltonen:2012it}
  T.~Aaltonen {\it et al.}  [CDF Collaboration],
  Phys.\ Rev.\ D {\bf 87} (2013) 092002
  [arXiv:1211.1003].

\bibitem{ArkaniHamed:1998rs}
  N.~Arkani-Hamed, S.~Dimopoulos and G.~R.~Dvali,
  Phys.\ Lett.\ B {\bf 429} (1998) 263
  [hep-ph/9803315];\\
  I.~Antoniadis, N.~Arkani-Hamed, S.~Dimopoulos and G.~R.~Dvali,
  Phys.\ Lett.\ B {\bf 436} (1998) 257
  [hep-ph/9804398].

\end{thebibliography}
\end{document}